\documentclass[12pt, preprint]{aastex}

\usepackage{natbib}

\newcommand{\Kvega}{K_{\rm vega}}

\newcommand{\Os}{\Omega_{\star}}
\newcommand{\Om}{\Omega_{\rm M}}
\newcommand{\Ol}{\Omega_{\Lambda}}
\newcommand{\Ob}{\Omega_b}
\def\avg#1{\left\langle#1\right\rangle}

\newcommand{\Lbox}{L_{\rm box}}

\newcommand{\Lam}{\Lambda}

\newcommand{\Del}{\Delta}
\newcommand{\hinv}{{h^{-1}}}
\newcommand{\mpc}{{\rm\,Mpc}}

\newcommand{\himpc}{\hinv{\rm\,Mpc}}
\newcommand{\hikpc}{\hinv{\rm\,kpc}}

\newcommand{\yr}{{\rm yr}}
\newcommand{\Msun}{{\rm M}_{\odot}}
\newcommand{\himsun}{\hinv{\Msun}}

\newcommand{\gtsim}{\gtrsim}
\newcommand{\Mstar}{M_\star}

\slugcomment{DRAFT VERSION}

\shorttitle{Massive Galaxies \& EROs at $z=1-3$}
\shortauthors{Nagamine et al.}

\begin{document}

\title{Massive galaxies \& EROs at $z=1-3$ in cosmological
hydrodynamic simulations: near-IR properties}

\author{Kentaro Nagamine\altaffilmark{1}, Renyue Cen\altaffilmark{2},  
Lars Hernquist\altaffilmark{3}, Jeremiah P. Ostriker\altaffilmark{2},\\  
\& Volker Springel\altaffilmark{4}}

\altaffiltext{1}{University of California, San Diego,
Center for Astrophysics \& Space Sciences, 9500 Gilman Dr.,
La Jolla, CA 92093-0424, U.S.A.}

\altaffiltext{2}{Princeton University Observatory, Princeton, NJ 08544,  
U.S.A.}

\altaffiltext{3}{Harvard-Smithsonian Center for Astrophysics,
60 Garden Street, Cambridge, MA 02138, U.S.A.}

\altaffiltext{4}{Max-Planck-Institut f\"{u}r Astrophysik,
         Karl-Schwarzschild-Stra\ss{}e 1, 85740 Garching bei
         M\"{u}nchen, Germany}


\begin{abstract}
Recent observations have revealed a population of red massive galaxies at 
high redshift which are challenging to explain in hierarchical galaxy 
formation models. We analyze this ``massive galaxy problem'' with two 
different types of hydrodynamic simulations -- Eulerian total variation 
diminishing (TVD) and smoothed particle hydrodynamics (SPH) --- of a 
concordance $\Lam$ cold dark matter ($\Lam$CDM) universe. We consider two 
separate but connected aspects of the problem posed by these extremely 
red objects (EROs): (1) the mass-scale of these galaxies, and (2) their 
red colors. We perform spectrophotometric analyses of simulated galaxies in 
$B, z, R, I,  J_s, K_s, K$ filters, and compare their near-infrared (IR) 
properties with observations at redshift $z=1-3$.  We find that the 
simulated galaxies brighter than the magnitude limit of $\Kvega=20$ mag 
have stellar masses $\Mstar\gtrsim 10^{11} \himsun$ and a number density 
of a few $\times 10^{-4}h^3 \mpc^{-3}$ at $z\sim 2$, in good agreement with 
the observed number density in the {\it K20} survey. Therefore, our 
hydrodynamic simulations do not exhibit the ``mass-scale problem''. The 
answer to the ``redness problem'' is less clear because of our poor 
knowledge of the amount of dust extinction in EROs and the uncertain 
fraction of star-forming EROs. However, our simulations can account for 
the observed comoving number density of $\sim 1\times 10^{-4}\,\mpc^{-3}$ 
at $z=1-2$ if we assume a uniform extinction of $E(B-V)=0.4$ for the 
entire population of simulated galaxies.
Upcoming observations of the thermal emission of dust in 24 $\mu$m by  
the {\it Spitzer Space Telescope} will help to better estimate the dust  
content of EROs at $z=1-3$, and thus to further constrain the star 
formation history of the Universe, and theoretical models of galaxy 
formation.
\end{abstract}



\section{Introduction}
\label{section:intro}

Multiband photometry including the near infrared (IR) band makes it 
possible to estimate the stellar mass of high redshift galaxies when
the observed photometric results are fitted
with artificial galaxy spectra generated 
by a population synthesis model. Working in the near-IR also allows one 
to construct a mass-selected sample, because near-IR band 
emission is less affected 
by dust extinction than at shorter wavelengths
and closely traces the total stellar mass. Using this 
technique, a number of recent observational  
studies have revealed a seemingly new population of very red, massive 
galaxies at redshift $z=1-3$ \citep[e.g.][]{Franx03, Rudnick03, 
Glazebrook04, McCarthy04a, Fontana04, Saracco04, Saracco05, Daddi04a, 
Daddi04b,Cimatti04}.

In this paper, we first give a brief review of the results of some of 
the major surveys that detected such a population of galaxies (see 
Section~\ref{sec:observations}).  All of these
recent observational studies, particularly at $z=1-3$, both in the UV  
and near-IR wavelengths, imply a range of novel tests for the hierarchical
structure formation theory.  We now face the important question as to  
whether this evidence for high-redshift massive galaxy formation is 
consistent with theoretical predictions based on the concordance 
$\Lam$CDM model. This question is sometimes called the 
``{\it massive galaxy problem}''.

In more detail, the ``{\it massive galaxy problem}'' that is posed by  
these observations can be divided into two separate but connected issues:  
(1) the ``{\it mass-scale problem}'', and 
(2) the ``{\it redness problem}''. The first aspect asks whether the 
hierarchical CDM model can produce a sufficient number of massive galaxies 
by $z=1-2$.  The second part of the problem, which seems to be more 
challenging for models, is whether there are enough very red, old, quiet, 
and massive galaxies. For example, \citet{Som04} compared the results of 
the {\it Great Observatories Origins Deep Survey} (GOODS) data and
the semi-analytic model of \citet{Som01}, and concluded that the 
semi-analytic model shows a deficit of both EROs and galaxies with 
$K<22$ at $z>1.5$, and that new or modified physics not yet accounted for 
in the semi-analytic models is needed to resolved this discrepancy.  
Therefore, it is of considerable interest whether hydrodynamic simulations 
of galaxy formation also suffer from the same problems.

In our first paper of the series \citep{Nachos1}, we argued that, based  
on two different types of hydrodynamic simulations (Eulerian TVD and SPH) 
and the theoretical model of \citet{Her03} (hereafter H\&S model), the 
predicted cosmic star formation rate (SFR) density peaks at $z\geq 5$, 
with a {\it higher} stellar mass density at $z=3$ than suggested by 
current observations, in contrast to some claims to the contrary.  This 
star formation history predicts that 70 (50, 30)\% of the total stellar 
mass at $z=0$ has already been formed by $z=1$ (2, 3). We also compared 
our results with those from the updated semi-analytic models of 
\citet{Som01}, \citet{Granato00}, and \citet{Menci02}, and found that 
our simulations and the H\&S model predicts an earlier peak of the SFR 
density and a faster build-up of stellar mass density compared to these 
semi-analytic models.

It is then interesting to examine what our simulations predict for the
properties of massive galaxies at $z\sim 2$. In our second paper
\citep{Nachos2}, we analyzed for this purpose the same set of  
hydrodynamic simulations, focusing on the UV properties of the most 
massive galaxies at the relevant epochs. Using the latest population 
synthesis model of \citet{BClib03}, we computed the spectra of simulated 
galaxies and performed a spectrophotometric analysis in the $U_n, G, R$ 
filter set.  We found that the simulated galaxies at $z=2$ satisfy the 
color-selection criteria proposed by \citet{Ade04} and \citet{Steidel04} 
when we assume a Calzetti extinction with $E(B-V) \sim 0.15$.  The total 
number density of simulated galaxies brighter than $R=25.5$ at $z=2$ was 
about $1\times 10^{-2}\,h^3\,{\rm Mpc}^{-3}$ for a uniform extinction of 
$E(B-V)=0.15$ (see Section~\ref{sec:discussion} for a discussion of the 
number density of UV and IR selected galaxies).  The most massive galaxies 
at $z=2$ have stellar masses $\geq 10^{11-12} \Msun$, and they typically 
have been continuously forming stars with a rate exceeding 
$30\,\Msun\,{\rm yr}^{-1}$ over a few Gyrs from $z=10$ to $z=2$, together 
with a significant number of starbursts reaching $1000\,\Msun\,\yr^{-1}$,
often lasting for a few tens of million years,   
superposed on the continuous component. TVD simulations indicated a more  
sporadic star formation history than the SPH simulations.  Those galaxies 
that appear to be red, passive systems at $z=2$ have completed the 
build-up of their stellar mass by $z\sim 3$, and have been quiet between 
$z=3$ and $z=2$.  We argued that our results imply that 
hierarchical galaxy formation {\it can} account for the massive galaxies 
at $z\geq 1$.

In this paper, we extend our analysis to the near-IR properties of massive
galaxies at $z=1-3$, with special focus on colors and stellar masses of
galaxies. The paper is organized as follows.  In
Section~\ref{sec:observations}, we give a brief summary of the current
observational situation with respect to EROs, and in
Section~\ref{sec:simulation} we describe the simulations we use.  In
Section~\ref{sec:method}, we review our method for computing spectra and
photometric magnitudes of simulated galaxies. We present our results for  
the near-IR properties of galaxies in Section~\ref{sec:results}. Finally, 
we summarize and discuss the implications of our work in 
Section~\ref{sec:discussion}.


\section{Observational data on massive galaxies and EROs} 
\label{sec:observations}

The Gemini Deep Deep Survey \citep[GDDS,][]{Abraham04} has obtained 225  
secure spectroscopic redshifts of the reddest and most luminous galaxies 
with $\Kvega<20.6$ mag.  These galaxies lie near the $I-K$ versus $I$
color-magnitude track mapped out by passively evolving galaxies in the
redshift interval $0.8 < z < 2$, probing the `redshift desert'. The 
infrared selection means that the GDDS is observing not only star-forming  
galaxies, as in most high-redshift galaxy surveys, but also quiescent 
evolved galaxies. About 25\% of their sample shows clear spectral 
signatures of evolved (pure old, or old + intermediate-age) stellar 
populations, 35\% shows features consistent with either a pure 
intermediate-age or a young + intermediate-age stellar population. 
About 29\% of the galaxies in the GDDS at $0.8<z<2$ are young starbursts 
with strong interstellar lines.  Using the GDDS data, \citet{Glazebrook04} 
estimated the stellar masses, and argued that there are a number of very 
massive, evolved red ($(I-K)_{\rm vega} > 4$) galaxies with stellar masses 
$M_\star > 10^{11}M_\odot$ at $z\sim 2$, which make a large contribution 
(30\%) to the stellar mass density in the Universe. \citet{McCarthy04a} 
estimated the ages of the red galaxies at $1.3<z<2.2$ in the GDDS data, 
and found that they have a median age of $1-3$ Gyrs with a history of 
a strong starburst phase ($300-500\,\Msun\,\yr^{-1}$) in the past.
These results suggest an early and rapid formation of massive galaxies  
at $z\gtrsim 1$.

Similarly, the recently completed {\it K20} survey 
\citep[e.g.,][]{Cimatti02a, Cimatti02c, Cimatti02b} identified 
a sample of over 500 
spectroscopic galaxies with $K_{s,\rm vega}<20$ with high 
spectroscopic completeness.  Among these, $\sim 30$ galaxies had 
spectroscopic redshifts $z>1.4$.  They also obtained $BVRIzJHK$ photometric 
data.  Using the {\it K20} data, \citet{Fontana04} 
demonstrated that there are 
galaxies with stellar masses $\Mstar>10^{11}\Msun$ at $z \simeq 2$. 
\citet{Cimatti02b} showed that some semi-analytic models of galaxy
formation underpredict the $K$ -band number counts compared to the 
{\it K20} survey data, and suggested a possible problem with current 
semi-analytic models. \citet{Daddi04a, Daddi04b} identified a significant 
population of $z=2$ galaxies with $K_{s,{\rm vega}}<20$ with high average 
star formation rates of $SFR\sim 200\,\Msun\,\yr^{-1}$ and median 
extinctions of $E(B-V)\sim 0.4$. These values are significantly higher than 
those of Lyman break galaxies' (LBGs') ($SFR\sim 40\,\Msun\,\yr^{-1}$ and 
$E(B-V)\sim 0.15$). \citet{Cimatti04} found 4 old, fully assembled, 
massive ($\Mstar>10^{11}\Msun$) spheroidal galaxies at $1.6<z<1.9$ in the 
{\it K20} data. The number density of such objects amounts to a few 
$\times 10^{-4}\,h^3\,\mpc^{-3}$.  In parallel to the {\it K20} survey,
\citet{Saracco05} discovered 7 bright ($17.8 <\Kvega< 18.4$) massive  
evolved galaxies at $0<z<1.7$ with $\Mstar=(3-6)\times 10^{11}\Msun$ in 
the Munich Near-IR Cluster Survey \citep[MUNICS: ][]{Drory01}.

The Faint InfraRed Extragalactic Survey \citep[FIRES,][]{Franx03} has  
revealed significant numbers of fairly bright galaxies at $z>2$ down to 
the magnitude limit of $K_s<24.4$ ($K_{s,{\rm vega}}=22.5$) selected by 
$(J_s-  K_s)_{\rm vega}>2.3$ colors. They named this population `distant 
red galaxies' (DRGs).  \citet{Forster04} and \citet{Dokkum04} performed 
near-IR spectroscopic analyses on $5-10$ DRGs, and found that they are 
more metal-rich ($\gtrsim$ solar), more massive ($\Mstar=1-5 \times 
10^{11}\Msun$) and older (ages of $1-2.5$ Gyr) than the $z=3$ LBGs, with 
extinctions of $A_V=2-3$ mags and extinction-corrected SFRs of 
$100-400\,\Msun\,\yr^{-1}$.  A plausible scenario that emerged from their 
study is that these DRGs are the descendants of LBGs at even higher 
redshifts, $z\gtrsim 5$. \citet{Rudnick03} estimated the stellar mass 
density from FIRES data at $z=0-3$ by combining the estimates of the 
rest-frame optical luminosity density and the mean cosmic mass-to-light
ratio. The FIRES group concluded that the red galaxies likely contribute
$\gtrsim 50\%$ of the stellar mass density in the Universe at $z\sim  
2.5$. In parallel to the FIRES, \citet{Saracco04} also discovered 3 
objects with $J_S-K_s>3$ at $z=2-3$ in the Hubble Deep Field South, and 
that these objects have already assembled $\Mstar\sim 10^{11}\Msun$ by 
then. Their results suggest that up to 40\% of the stellar mass content 
of bright ($L>L^*$) local early type galaxies was already in place at 
$z>2.5$.

Other works on the global stellar mass density in the Universe in the  
redshift range of $0\le z \le 3$ include those by \citet{Brinch},
\citet{Cole}, \citet{Cohen}, \citet{Dick03a}, and \citet{Fontana03}.
These observational estimates constrain 
the evolution of the stellar mass density $\Os$ as a function of redshift 
or cosmic time. By comparing observational data and semi-analytic models 
of galaxy formation \citep[e.g.][]{Kau99, Som01, Cole00}, some authors 
have argued that the hierarchical structure formation theory may have 
difficulty in accounting for sufficient early star formation 
\citep[e.g.,][]{Poli, Fontana03, Dick03a}.

In addition to these recent findings on the red massive galaxies,
\citet{Ade04} and \citet{Steidel04} have introduced new techniques for
exploring the so-called `redshift desert', making it possible to 
efficiently identify 
a large number of galaxies that are bright in the ultra-violet (UV) 
wavelengths  
with the help of a color selection criteria in the color-color 
plane of $U_n -G$ versus $G-R$. In this technique, galaxies at $z=2 - 2.5$ 
are located photometrically from the mild drop in the $U_n$ filter owing 
to the Ly-$\alpha$ forest opacity, and galaxies at $z=1.5-2$ are recognized  
from the lack of a break in their observed-frame optical spectra.  The 
large sample of UV bright galaxies identified by these authors at $z\sim 2$ 
makes it now possible to study galaxy formation and evolution for over 
10 Gyrs of cosmic time, from redshift $z=3$ to $z=0$, without a 
significant gap.  We note that the epoch around $z\sim 2$ is particularly 
important for understanding galaxy evolution because the number density of 
quasi-stellar objects (QSOs) peaks at $z=1-2$ 
\citep[e.g.,][]{Schmidt68, Schmidt95, Fan01b, Barger05} and the UV luminosity 
density began to decline by about an order of magnitude from $z\sim 2$ to 
$z=0$ \citep[e.g.,][]{Lilly96, Connolly97, Sawicki97, Treyer98, Pas98, 
Cowie99}.

At the same time, there has been mounting evidence for high redshift  
galaxy formation including the discovery of Extremely Red Objects (EROs, 
often defined as $(I-K)_{\rm vega}>4$ or $(R-K)_{\rm vega}>5$) at $z\ge 1$
\citep[e.g.,][]{Elston88, McCarthy92, Hu94, Smail02, Cimatti03,  
Saracco03, McCarthy04b}, sub-millimeter galaxies at $z\ge 2$  
\citep[e.g.,][]{Smail97, Chapman03}, Lyman break galaxies (LBGs) at 
$z\geq 3$ \citep[e.g.,][]{Steidel99, Iwata, Ouchi04}, and Lyman-$\alpha$  
emitters at $z\geq 4$ \citep[e.g.,][]{Hu99, Rhoads01, Taniguchi, Kodaira, 
Ouchi03}.

We will give further details on the observations of EROs in the relevant 
subsequent sections.


\section{Simulations}
\label{sec:simulation}

In this section, we describe the two different types of cosmological
hydrodynamic simulations that we use in this paper.  Both approaches 
include ``standard'' physics such as radiative cooling/heating, star 
formation, and supernova (SN) feedback, although the details of the models 
and the parameter choices vary considerably.

One set of simulations was performed using an Eulerian approach,
which employed a particle-mesh method for the gravity and the total
variation diminishing (TVD) method \citep{Ryu93} for the hydrodynamics,
both with a fixed mesh. The treatment of the radiative cooling and
heating is described in \citet{Cen92}. The structure of
the code is similar to that of \citet{CO92, CO93}, but it has
significantly improved over the years with additional input physics.
It has been used for a variety of studies, including the evolution
of the intergalactic medium \citep{CO94, CO99a, CO99b, CNO04},
damped Lyman-$\alpha$ absorbers \citep{Cen03}, and galaxy formation
\citep*[e.g.][]{CO00, Nag01a, Nag01b, Nag02}.

The other set of simulations employs the Lagrangian smoothed particle
hydrodynamics (SPH) technique.  We used an updated version of the code
{\small{GADGET}} \citep{Gadget}, based on an `entropy conserving'  
formulation \citep{SH02} of SPH that alleviates problems with 
energy/entropy conservation \citep[e.g.][]{Her93} and numerical 
overcooling. The code also includes a subresolution multiphase model 
for the interstellar medium, a phenomenological model for feedback by 
galactic winds \citep{SH03a}, and the impact of a uniform ionizing 
radiation field \citep{KWH96, Dave99}.  These simulations have been used 
previously to study the evolution of the cosmic SFR \citep{SH03b, Nachos1}, 
damped Lyman-$\alpha$ absorbers \citep*{NSH04a, NSH04b}, Lyman-break 
galaxies \citep*{NSHM}, disk formation \citep{Robertson04}, emission 
from the intergalactic medium \citep{Fetal03,Fetal04a, Fetal04b, Fetal04c, 
Fetal04d, Zetal04}, and the detectability of high redshift galaxies 
\citep{Bart04, Fetal04e}.

The cosmological parameters adopted in the simulations are consistent with
recent observational determinations \citep[e.g.][]{Spergel03}, as 
summarized in Table~\ref{table:simulation}, where we list the most 
important numerical parameters of our primary runs. While there are many 
similarities in the physical treatment between the two approaches (TVD and 
SPH), they differ in their relative resolution as a function of density. 
Broadly speaking, the TVD simulations tend to have better mass resolution 
in low density regions, while the SPH simulations tend to have better 
spatial resolution in high-density regions. In this sense, the two 
approaches can be viewed as complementary, and results found in common 
can be expected to be robust.


\section{Analysis Method}
\label{sec:method}

In this section, we briefly describe our spectrophotometric analysis method,
which is based on the same techniques for identifying the galaxies in our
simulations and computing their spectra as in \citet{Nachos1}.

We use the population synthesis model by \citet{BClib03}, assuming a
\citet{Chab03b} initial mass function (IMF) within a mass range of  
$[0.1, 100]\,\Msun$, as recommended by \citet{BClib03}.  Spectral 
properties obtained with this IMF are very similar to those obtained using 
the \citet{Kroupa01} IMF, but the \citet{Chab03b} IMF provides a better 
fit to counts of low-mass stars and brown dwarfs in the Galactic disk 
\citep{Chab03a}. We use the high resolution version of the spectral 
library of \citet{BClib03} which contains 221 spectra describing the 
spectral evolution of a ``simple stellar population'' from $t=0$ to $t=20$~Gyr 
over 6900 wavelength points in the range of 91~\AA - 160~$\mu$m.  Based on 
the stellar mass, metallicity, and the formation time of each star particle 
in the simulations, we compute the spectrum of each star particle treating 
it as a simple stellar population. We then later co-add them to obtain the 
spectra of simulated galaxies, each typically containing hundreds to 
thousands of star particles.

Once the intrinsic spectrum is computed, we apply the \citet{Calzetti00}
extinction law with different values of $E(B-V)=0.0$, $0.15$, $0.4$,  
$0.75$, $1.0$, in order to investigate the impact of internal extinction 
within the galaxies. Because the median extinction of the {\it K20} 
galaxies is $E(B-V)=0.4$, we take this value as our fiducial value in the  
following. Note that the analysis presented in \citet{Nachos1} was limited 
to $E(B-V)\leq 0.3$. Using the spectra computed in this manner, we then 
derive the rest-frame colors and luminosity functions of the simulated 
galaxies.

To obtain the spectra in the observed frame, we redshift the spectra and 
apply absorption by the intergalactic medium (IGM), following the 
prescription of \citet{Madau95}.  Once the redshifted spectra in the 
observed frame are obtained, we convolve them with different filter 
functions, including $U_n, G, R$ \citep{Steidel93}, standard Johnson bands, 
and SDSS bands as well as $J_s$ and $K_s$-bands, and compute the magnitudes 
in the AB system.  (See \citet{Night05} for the details of this procedure.)
All the magnitudes are presented in the AB system unless otherwise 
indicated.  Where a conversion between AB and Vega system is necessary, 
we use the following relationships and explicitly mention which system is 
being used in the subscript: $R_{AB} = R_{\rm vega} + 0.27$, 
$I_{AB}=I_{\rm vega}+0.50$, $J_{s,AB}=J_{s,{\rm vega}}+0.92$, 
$K_{\rm AB} = K_{\rm vega} + 1.88$, therefore 
$(R-K)_{AB}=(R-K)_{\rm vega} - 1.6$, $(I-K)_{AB}=(I-K)_{\rm vega} -  1.4$, 
and $(J_s-K_s)_{AB}=(J_s-K_s)_{\rm vega}-0.96$.  These values were obtained  
by computing the AB magnitude of the Vega star, using the \citet{Kurucz92}
model spectra for Vega included in the population synthesis package by
\citet{BClib03}, with the normalization of $f_\lambda= 3.44\times  10^{-9}$ 
erg cm$^{-2}$ s$^{-1}$ \AA$^{-1}$ at 5556 \AA\ \citep{Hayes85}.  For example, 
the color-cut of $(I-K)_{\rm vega}>4$, $(R-K_s)_{\rm vega}>5$ (for the  
EROs), $(J_s-K_s)_{\rm vega}>2.3$ (for the DRGs) corresponds to $I-K>2.6$,
$R-K_s>3.4$, $J_s-K_s>1.34$ in the AB system.


\section{Results}
\label{sec:results}

\subsection{$K_s$ magnitude and stellar mass}
\label{sec:kmag_mstar}

\begin{figure}
\epsscale{1.0}
\plotone{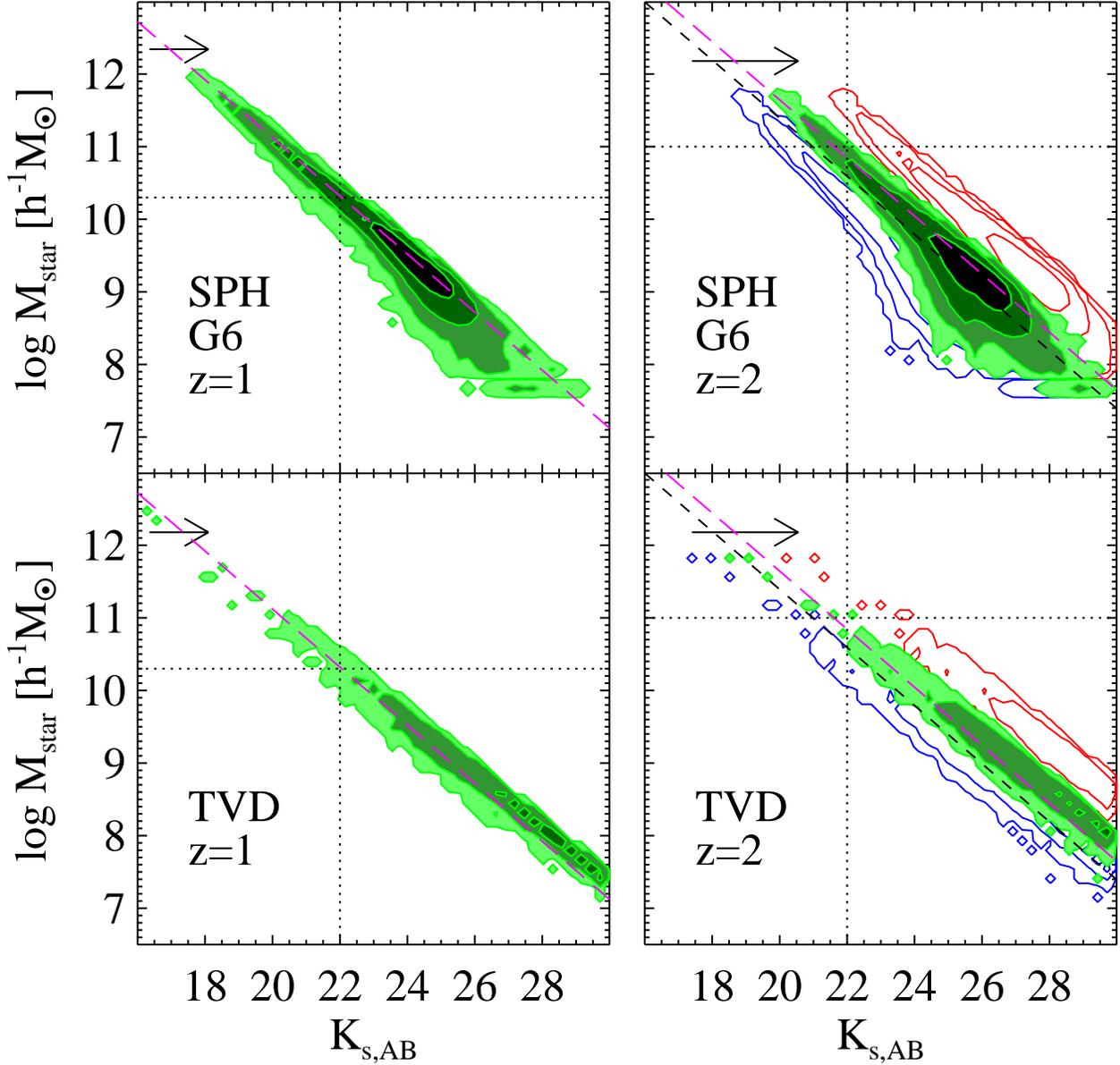}
\caption{$K$-band magnitude versus stellar mass of simulated galaxies at 
$z=2$. The two dashed lines in the right column panels correspond to the 
relation found by \citet{Daddi04b} from the {\it K20} survey observational 
data (see text for details).  In the right column panels, three sets of 
distributions are shown in blue, green, and red colors, corresponding to 
the extinction values $E(B-V)=0.0$, 0.4, and 1.0, respectively, with an 
arrow indicating the increasing extinction from 0.0 to 1.0. 
In the left column panels, only the case for $E(B-V)=0.4$ for $z=1$ is 
shown and the magenta dashed line is a similar relation to those in the 
right column panels with different normalization (see text for the exact
value). The vertical dotted line indicates the magnitude limit of 
$K_S=22$, and the horizontal dotted line indicates the mass-scale of 
$\Mstar=10^{10.3}\himsun$ for $z=1$ (left column) and $10^{11}\himsun$ for 
$z=2$ (right column).  }
\label{f1.eps}
\end{figure}

In Figure~\ref{f1.eps}, we plot the $K_s$ magnitude versus stellar mass  
for galaxies at $z=1$ and $2$, for SPH as well as TVD runs.  In the right 
column panels, three sets of distributions are shown in blue, green, and red 
colors, respectively, corresponding to the extinction values $E(B-V)=0.0$, 
$0.4$, and $1.0$, with an arrow indicating the increasing extinction from 
0.0 to 1.0.  The two dashed lines in the right column panels for $z=2$ 
correspond to the relation found by \citet{Fontana04} and \citet{Daddi04b}:
$\log(\Mstar/10^{11}\Msun)=-0.4(K-K^{11})$, where $K_{AB}^{11}=21.37$  
and 22.01 depends on the method used for the estimate.  
The good agreement between the simulation results and the observational 
relation is very encouraging. Slight deviations of the simulation results 
from the lines can be attributed to the scatter in the extinction value  
which is not taken into account in our theoretical calculations.
In the left column panels, only the case for $E(B-V)=0.4$ is shown and 
the magenta dashed line is for $K_{AB}^{11}=20.7$.  

Note that the scatter in the distribution of the simulated galaxies is  
much smaller compared with the same diagram as a function of $R$-band  
magnitude \citep[Fig.~3 of][]{Nachos2}. This is because the $K$-band 
magnitude traces the stellar mass better than the $R$-band magnitude 
which probes the rest-frame UV wavelengths at $z=2-3$.

Also from this figure, we see that the magnitude limit of $K_s=22$  
(i.e., $K_{s, {\rm vega}}= 20$ mag) roughly corresponds to the stellar mass  
$\Mstar \sim 10^{10.3}\himsun$ (at $z=1$) and $10^{11}\himsun$ (at $z=2$), 
as indicated by the vertical dotted line.  As was shown in panel (a) of 
Fig.~4 in \citet{Nachos2}, the corresponding number density at $z=2$ 
above the limit of $\Mstar=10^{11}\himsun$ is $n(\Mstar>10^{11}\himsun) = 
3.5\times 10^{-4}\,h^3\,\mpc^{-3}$ for the SPH G6 run, and $6.6\times 
10^{-4}\,h^3\,\mpc^{-3}$ for the TVD run.  The number density at $z=2$ 
above the magnitude limit of $K_s=22$ is 
$n(K_s<22) = 4.5\times 10^{-4}\,h^3\,\mpc^{-3}$ for the SPH G6 run, and 
$5.6\times 10^{-4}\,h^3\,\mpc^{-3}$ for the TVD run.
The agreement between the two runs is reasonable when effects owing to the
differences in boxsize, simulation methodology, and cosmic variance are 
taken into account.  

We have repeated the same exercise at $z=3$. To summarize, we find that 
the following values of $K_{AB}^{11}$ describes the relation between the 
$K_s$ magnitude and the stellar mass of simulated galaxies: 
$K_{AB}^{11}= 20.7$ (for z=1), 22.01 (for $z=2$, as given above by the K20 
survey), and 23.0 (for z=3).


\subsection{Color -- magnitude diagrams}

\subsubsection{$I-K$ versus $I$ at $z=1$}

\begin{figure*}
\begin{center}
\resizebox{16.0cm}{!}{\includegraphics{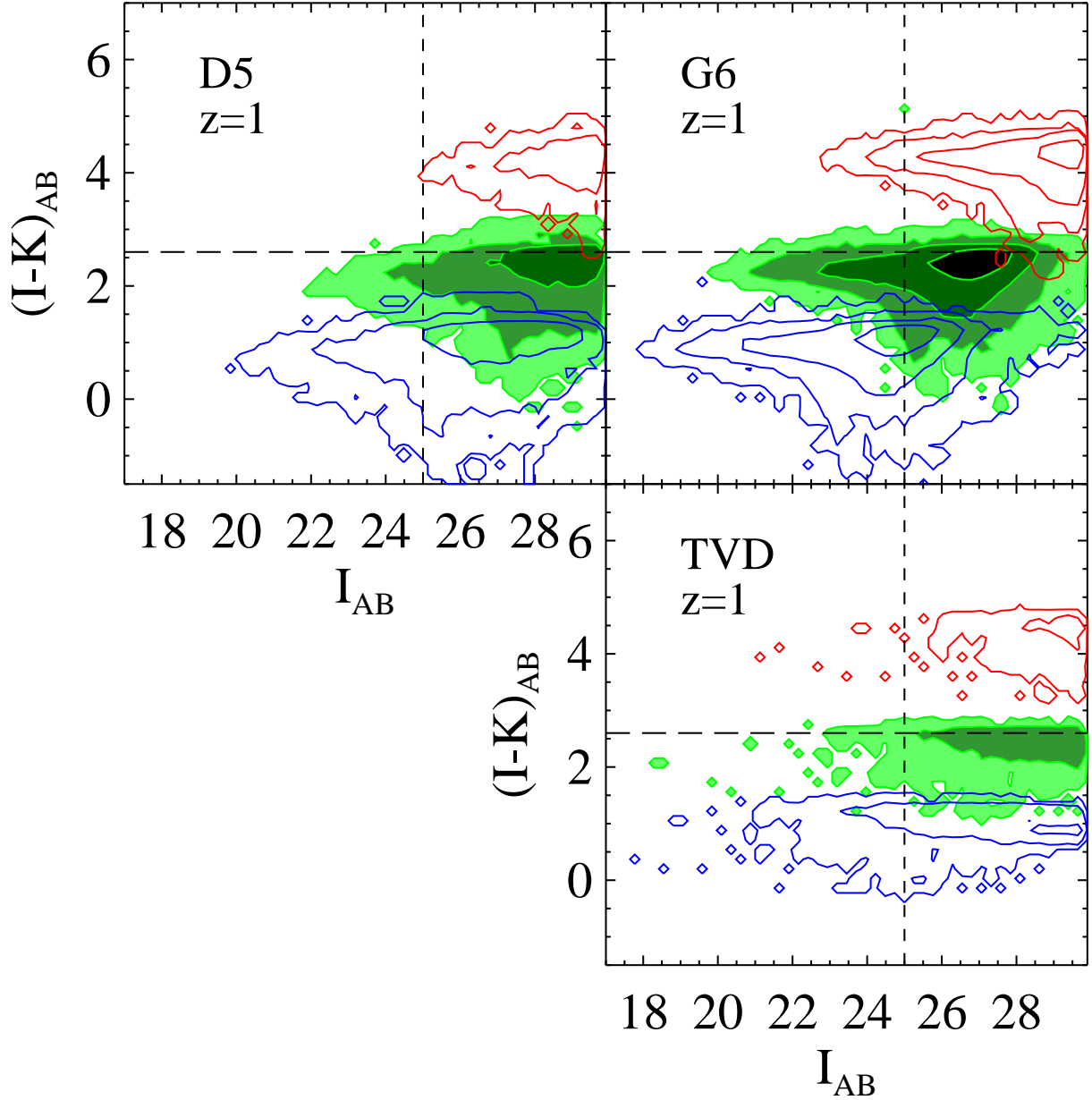}}
\vspace{0.7cm}
\caption{Color-magnitude diagram in the plane of $(I-K)_{AB}$ color 
versus $I_{AB}$-band magnitude for the SPH and TVD runs at $z=1$, 
corresponding to those of the GDDS \citep{Abraham04}.  The three 
different sets of distributions represent different extinction values: 
$E(B-V)=0.15$ (blue), 0.4 (green), and 1.0 (red). The magnitude limit 
of $I_{AB}=25$ of GDDS is indicated by the vertical dashed line. 
The color-cut $(I-K)_{AB}>2.6$ (i.e., $(I-K)_{\rm vega}>4$) for the ERO 
selection is also indicated as a long-dashed line.  }
\label{f2.eps}
\end{center}
\end{figure*}

Figure~\ref{f2.eps} shows the color-magnitude diagram of $I-K$ versus  
$K$-band magnitude at $z=1$, both for SPH and TVD simulations.  This diagram
corresponds to those presented by the GDDS group, e.g., Fig. 6 of
\citet{Abraham04} for galaxies at $z\sim 1$.  The three different
distributions represent different extinction values of $E(B-V)=0.0$ (blue),
0.4 (green), and 1.0 (red). The magnitude limit of $I_{AB}=25$ (i.e.,  
$I_{\rm vega}=24.5$) of GDDS is indicated by the vertical dashed line.
The color-cut $I-K>2.6$ (i.e., $(I-K)_{\rm vega}>4$) for the ERO selection  
is also indicated as a long-dashed line.  Our simulations suggest 
extinction values of $E(B-V)\gtrsim 0.4$ for the EROs with $I-K>2.6$ at 
$z\sim 1$.  Assuming $E(B-V)=0.4$ uniformly for the entire population, 
the corresponding number density of EROs at $z=1$ ($I<25$ and $I-K>2.6$) 
within the magnitude limit is $2.3\times 10^{-4}\,h^3\,\mpc^{-3}$ for the 
SPH D5 run, $3.2\times 10^{-4}\,h^3\,\mpc^{-3}$ for the SPH G6 run, and 
$2.5\times 10^{-3}\,h^3\,\mpc^{-3}$ for the TVD run.

For comparison, the GDDS sample in \citet{McCarthy04a} with 
$(I-K)_{\rm vega}>4$ (16 objects) contributes $n = (3.4 _{-1.2}^{+1.3}) 
\times 10^{-4}\,h^3\,\mpc^{-3}$ at $1.3<z<2$, in good agreement with 
our SPH results. The TVD run predicts a higher number density for the 
$I-K$ selected EROs at $z=1$ compared to the GDDS data.


\subsubsection{$R-K_s$ versus $K_s$ at $z=1-3$}
\label{sec:rk}

\begin{figure}
\epsscale{1.0}
\plotone{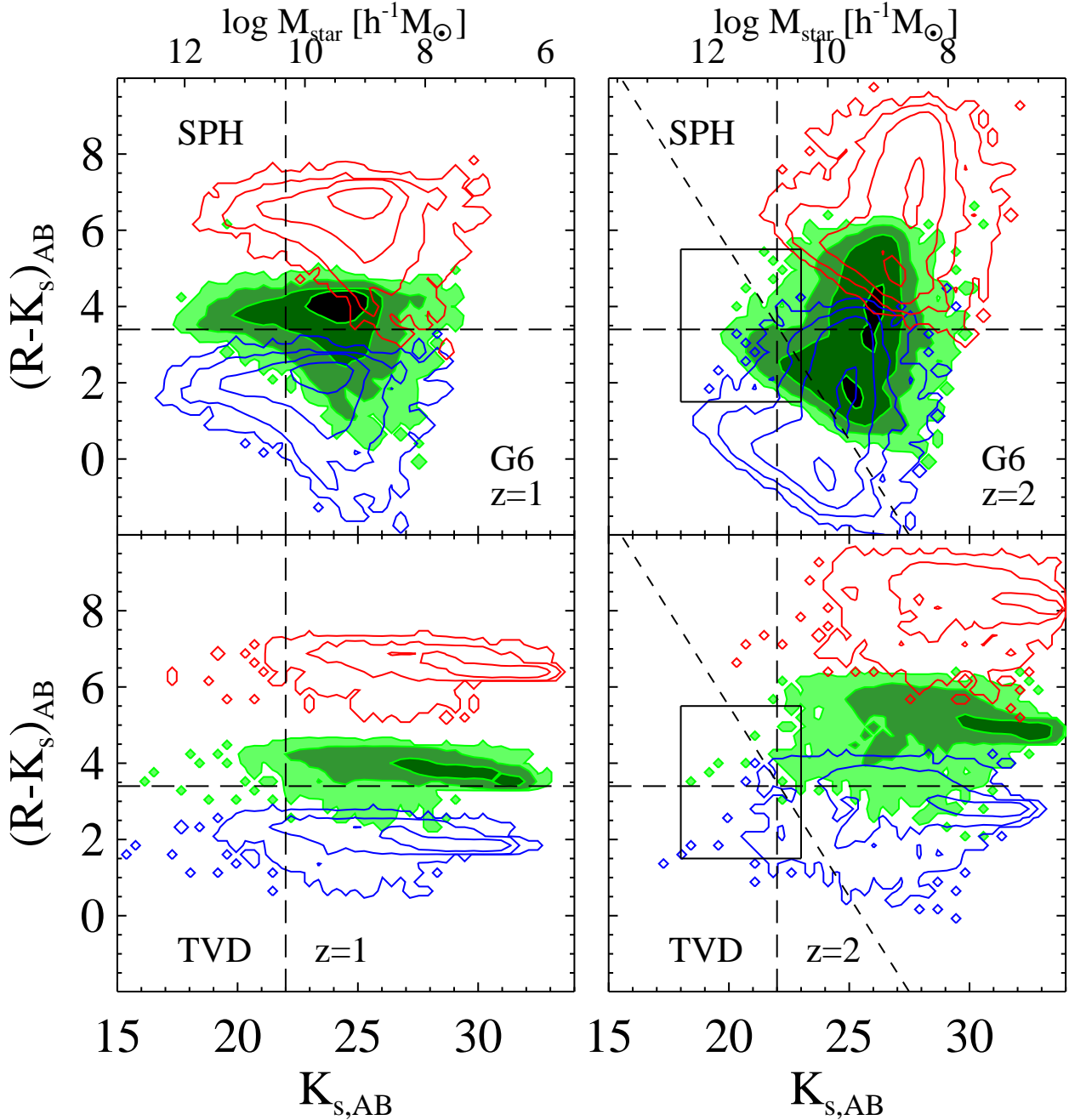}
\caption{Color-magnitude diagram in the plane of $R-K_s$ color versus
$K_s$-band magnitude for the SPH G6 run (upper row) and the TVD N864L22 
run (bottom row) at $z=1$ (left column) and $z=2$ (right column). 
The top axes indicate the corresponding mass-scale obtained 
by the relation described in Section~\ref{sec:kmag_mstar}. The three 
different sets of distributions in each panel correspond to extinction  
values $E(B-V)=0.0$ (blue), 0.4 (green), and 1.0 (red). 
The magnitude limit of $K_s<22$ and the color-cut of $R-K_s>3.4$ 
(i.e., $(R-K)_{\rm vega}>5$) for the ERO selection is indicated by the 
vertical and horizontal long-dashed line, respectively.  The square box 
shown in the right column panels encompasses the same region of space as 
Fig.~9 of \citet{Steidel04}, for comparison with a UV selected sample.  
The slanted short-dashed line in the right column panels indicates the 
constant magnitude limit of $R=25.5$.  }
\label{f3.eps}
\end{figure}

In Figure~\ref{f3.eps}, we show the color-magnitude diagram of $R-K_s$  
versus $K_s$-band magnitude at $z=1$ and $z=2$, both for the SPH G6 run 
and TVD runs. The top axes indicate the corresponding mass-scale obtained 
by the relation described in Section~\ref{sec:kmag_mstar}.  
In each panel, three different distributions are shown for extinctions 
$E(B-V)=0.0$ (blue), 0.4 (green), and 1.0 (red). The magnitude limit of 
$K_s=22$ and the color-cut of $R-K_s>3.4$ (i.e., $(R-K_s)_{\rm vega}>5$) for 
the ERO selection is indicated by the long-dashed lines. Similarly to 
Fig.~\ref{f2.eps}, our simulations suggest that extinction of 
$E(B-V)\gtrsim 0.4$ is needed to account for the EROs with $R-K_s>3.4$ at 
$z\sim 2$.

Assuming a uniform extinction of $E(B-V)=0.4$ for the entire population of
simulated galaxies, the corresponding number density of EROs at $z=1-2$  
that satisfy $K_s<22$ is summarized in Table~\ref{table:ERO}.  If we change  
the magnitude limit to $R<25.5$ instead (with $R-K_s>3.4$), we then obtain
$5.0\times 10^{-5}\,h^3\,\mpc^{-3}$ (SPH G6 run) and $3.8\times
10^{-4}\,h^3\,\mpc^{-3}$ (TVD run) at $z=2$.  At $z=3$ there are only a
handful of galaxies that satisfy $R-K_s>3.4$ and $R<25.5$, and the  
numbers are clearly affected by the small statistics: 
$5.0\times 10^{-6}\,h^3\,\mpc^{-3}$ for the SPH G6 run (5 objects in a 
$\Lbox=100\himpc$ box), and a null result for the TVD run.

\citet{Mous04} estimated the space density of EROs using the GOODS data,
finding $1.87\times 10^{-3}\,h^3\,\mpc^{-3}$ for the EROs with $K_s<22$,
$R-K_s>3.35$, and a median redshift of $z_{\rm med}=1.2$.  The simulated
number density of EROs at $z=1$ listed in Table~\ref{table:ERO} is 
slightly higher than their estimate, but they are reasonably close to 
each other considering the uncertainty in the distribution of the 
extinction parameter. \citet{Cimatti02a} estimated the number density of 
EROs with $\Kvega<19.2$ and $(R-K_s)_{\rm vega}>5$, and found 
$6.3\times 10^{-4}\,h^3\,\mpc^{-3}$, whereas we find somewhat higher values 
of $1.4\times 10^{-3}\,h^3\mpc^{-3}$ (SPH G6 run) and 
$2.0\times 10^{-3}\,h^3\mpc^{-3}$ (TVD run) for the same magnitude
limit and a uniform extinction of $E(B-V)=0.4$ at $z=1$. \citet{Vaisanen04}
also estimated the ERO number density to be $\approx 5.8\times
10^{-5}\,h^3\,\mpc^{-3}$ for the EROs with $\Kvega<17.5$, using the  
European Large Area ISO Survey (ELAIS) data.  For the same magnitude limit 
and a uniform extinction of $E(B-V)=0.4$, we find $1.5\times 
10^{-4}\,h^3\,\mpc^{-3}$ (SPH G6 run) and $5.6\times 10^{-4}\,h^3\,\mpc^{-3}$
(TVD run) at $z=1$.  \citet{Saracco05} obtained the comoving number  
density of $1.6\times 10^{-4}\,h^3\,\mpc^{-3}$ for the EROs with 
$17.8<K<18.4$ and $(R-K)_{\rm vega}>5.3$ in the MUNICS data. In comparison, 
we find $3.3\times 10^{-4}\,h^3\,\mpc^{-3}$ (SPH G6 run) and $1.9\times  
10^{-4}\,h^3\,\mpc^{-3}$ (TVD run) at $z=1$ for the same magnitude range 
and color-cut.

The number density of $R-K_s$ selected simulated galaxies (with $K_s<22$) 
is higher than that of $I-K$ selected (with $I<25$) galaxies. This is  
consistent with the finding of \citet{Vaisanen04} that the $R-K$ selected 
EROs have higher number counts than the $I-K$ selected EROs.

In the panels of the right column for $z=2$ in Figure~\ref{f3.eps}, 
black square boxes that  
encompass the same region of space as Fig.~9 of \citet{Steidel04} are 
shown.  The dashed line indicates the constant magnitude limit of $R=25.5$,  
therefore, the observed data lie inside the box below this dashed line.  
We see that the simulated galaxies occupy a similar region in the 
color-magnitude plane as Steidel's sample at $z=2$, suggesting some 
overlap between the near-IR selected sample and that of \citet{Steidel04}.  
The figure also suggests that the UV selected sample does not contain 
highly extincted galaxies with $E(B-V)\geq 0.75$.


\subsubsection{$J_s-K_s$ versus $K_s$ at $z=2-3$}
\label{sec:jk}

\begin{figure}
\epsscale{1.0}
\plotone{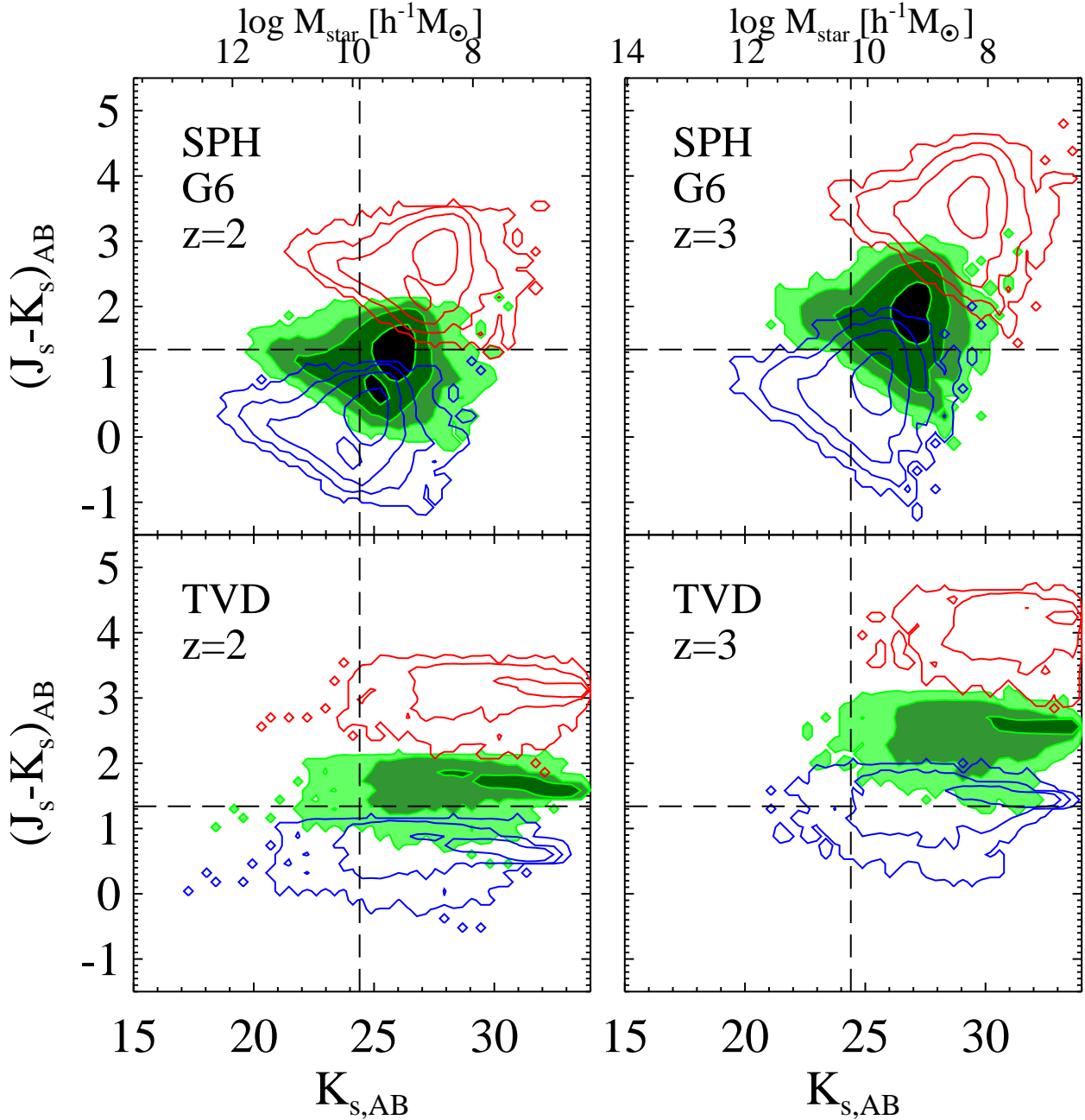}
\caption{Color-magnitude diagram in the plane of $J_s-K_s$ color versus 
$K_s$-band magnitude for the SPH G6 run and the TVD run at $z=2$ and 3. 
The vertical dashed line indicates the magnitude limit of $K_s<24.4$
(i.e., $\Kvega\lesssim 22.5$), and the color-cut of $J_s-K_s>1.34$ 
(i.e., $(J_s-K_s)_{\rm vega}>2.3$) is also shown as the horizontal  
dashed line.  The four different colors represent different extinction 
values: $E(B-V)=0.0$ (blue), 0.4 (green), and 1.0 (red).  }
\label{f4.eps}
\end{figure}

Figure~\ref{f4.eps} shows the color-magnitude diagram of $J_s-K_s$ color
versus $K_s$-band magnitude, at redshifts $z=2$ and 3, both for SPH G6  
run and TVD runs. This diagram has been used by the FIRES group 
\citep[e.g.][]{Franx03, Dokkum04, Forster04}.  The top axes indicate the 
corresponding mass-scale obtained by the relation described in 
Section~\ref{sec:kmag_mstar}. The three different distributions in each 
panel represent extinction values of $E(B-V)=0.0$ (blue), 0.4 
(green), and 1.0 (red).  The vertical short-dashed lines indicate the 
magnitude limit of $K_s<24.4$ (i.e., $K_{s,{\rm vega}}\lesssim 22.5$), 
and the color-cut of $J_s-K_s>1.34$ (i.e., $(J_s-K_s)_{\rm vega}>2.3$) is 
shown as a horizontal long-dashed line.

Similar to Fig.~\ref{f2.eps}, our simulations suggest extinction values  
of $E(B-V)\gtrsim 0.4$ for galaxies with $J_s-K_s>1.34$ at $z\sim 2$.   
Assuming $E(B-V)=0.4$ uniformly for the entire $z=2$ sample, the 
corresponding number densities for the above magnitude limit and the 
color-cut are $1.4\times 10^{-3}\,h^3\,\mpc^{-3}$ for the SPH G6 run and 
$7.4\times 10^{-3}\,h^3\,\mpc^{-3}$ for the TVD run.  Similarly at $z=3$,  
$1.3\times 10^{-3}\,h^3\,\mpc^{-3}$ for the SPH G6 run and $1.5\times
10^{-3}\,h^3\,\mpc^{-3}$ for the TVD run.

Because the magnitude limit ($K_{s,AB}=24.4$) of FIRES is a few magnitudes
deeper than that of the GDDS and the {\it K20} survey, galaxies with  
stronger extinction ($E(B-V)\gtrsim 0.5$) can be sampled better, as seen 
in Fig.~\ref{f4.eps}. This is in accordance with the fact that the amount  
of extinction estimated from the FIRES data is in the range $A_V=1-3$  
mags, which corresponds to $E(B-V)=0.25-0.74$ for the \citet{Calzetti00} 
extinction law.  Comparison of Fig.~\ref{f3.eps} and Fig.~\ref{f4.eps} 
suggests that there would be a significant overlap between EROs 
(defined as $(R-K_S)_{\rm vega}>5$) and DRGs (defined as 
$(J_S-K_s)_{\rm vega}>2.3$), but the overlap may not be complete.


\subsection{$BzK_s$ diagram at $z=2$}
\label{sec:bzk}

Recently, \citet{Daddi04b} devised a new color-selection technique in  
order to separate star-forming galaxies and quiescent old galaxies for the  
$K_s$-band bright galaxies. They defined $BzK_s \equiv (z-K_s)_{AB} - 
(B-z)_{AB}$,  and demonstrated, using the {\it K20} data, that the line 
of $BzK_s = -0.2$  in the color-color plane of $B-z$ versus $z-K_s$ 
('$BzK_s$ diagram') separates star-forming and quiescent old galaxies 
without star formation (indicated as `dead \& red') at $z>1.4$ quite 
nicely. The star-forming galaxies can hence be isolated in a  $BzK_s$ 
diagram by $BzK>-0.2$, (upper left part of the plot) and the old galaxies 
by the criteria of $BzK_s<-0.2$ combined with $z-K_s>2.5$ (upper right 
corner of the plot).  The $z-K_s>2.5$ criteria is similar to the commonly 
employed color-cut $R-K_s>5$ for selecting EROs. A convenient
feature of this 
color separation technique is that the the extinction vector is almost 
parallel to the line of $BzK_s=-0.2$, therefore, the separation of the 
two populations is not strongly dependent on the amount of dust reddening.

\begin{figure}
\epsscale{1.0}
\plotone{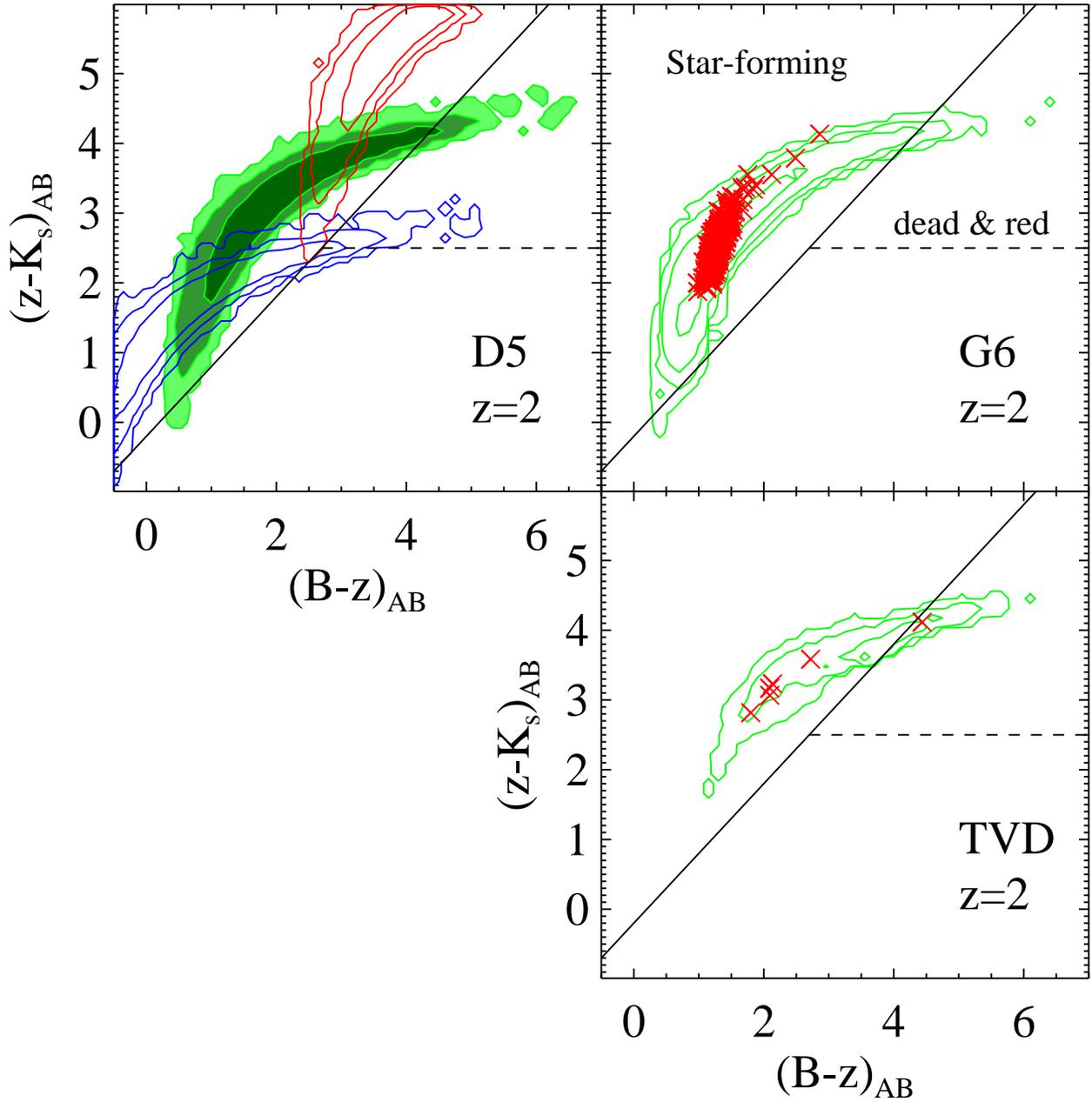}
\caption{`BzK' diagram at $z=2$ for the SPH and TVD runs. The lines 
$BzK_s = -0.2$ and $z-K_s=2.5$ are shown by the solid and the dashed lines, 
respectively.  In the top left panel for the SPH D5 run, we show three 
different sets of distributions corresponding to $E(B-V)=0.0$ (blue),
0.4 (green), and 1.0 (red). The red crosses overplotted in the SPH G6 
and the TVD panels are the galaxies that are brighter than 
$K_s=22$ (i.e., $K_{s,{\rm vega}}<20$).  }
\label{f5.eps}
\end{figure}

Figure~\ref{f5.eps} shows the $BzK_s$ diagram of simulated galaxies at  
$z=2$, both for SPH and TVD runs. The lines $BzK_s = -0.2$ and $z-K_s=2.5$ 
are shown by the solid and the dashed lines, respectively.  In the top 
left panel, for the SPH D5 run, we show three different distributions 
corresponding to $E(B-V)=0.0$ (blue), 0.4 (green), and 1.0 (red). 
In the panels for the SPH G6 and TVD runs, only the case for 
$E(B-V)=0.4$ is shown. The red crosses overplotted in the SPH G6 and the 
TVD panels are the galaxies that are brighter than $K_s=22$ (i.e., 
$K_{s,{\rm vega}}\lesssim 20$). In the SPH G6 run, all the red
crosses are in the star-forming region, but in the TVD simulation there  
is one galaxy that satisfies $BzK_s<-0.2$ and $z-K_s>2.5$.

In the SPH G6 run, the corresponding number density of galaxies at  
$z=2$ with $K_s<22$ and $z-K_s>2.5$ is $2.2\times 10^{-4}h^3\mpc^{-3}$.  
For the TVD run, we find $5.6\times 10^{-4}h^3\mpc^{-3}$, but if we also 
require $BzK_s<-0.2$ we have $1.0\times 10^{-4}h^3\mpc^{-3}$.  These number 
densities are roughly consistent with the observed number density of 
$\sim 10^{-4} \mpc^{-3}$ for galaxies with $K_{S,{\rm vega}}<20$ and 
$E(B-V)\sim 0.4$ by \citet{Daddi04b}. See Section~\ref{sec:discussion} 
for further discussion on the comparison of our results with observations.

\begin{figure}
\epsscale{1.0}
\plotone{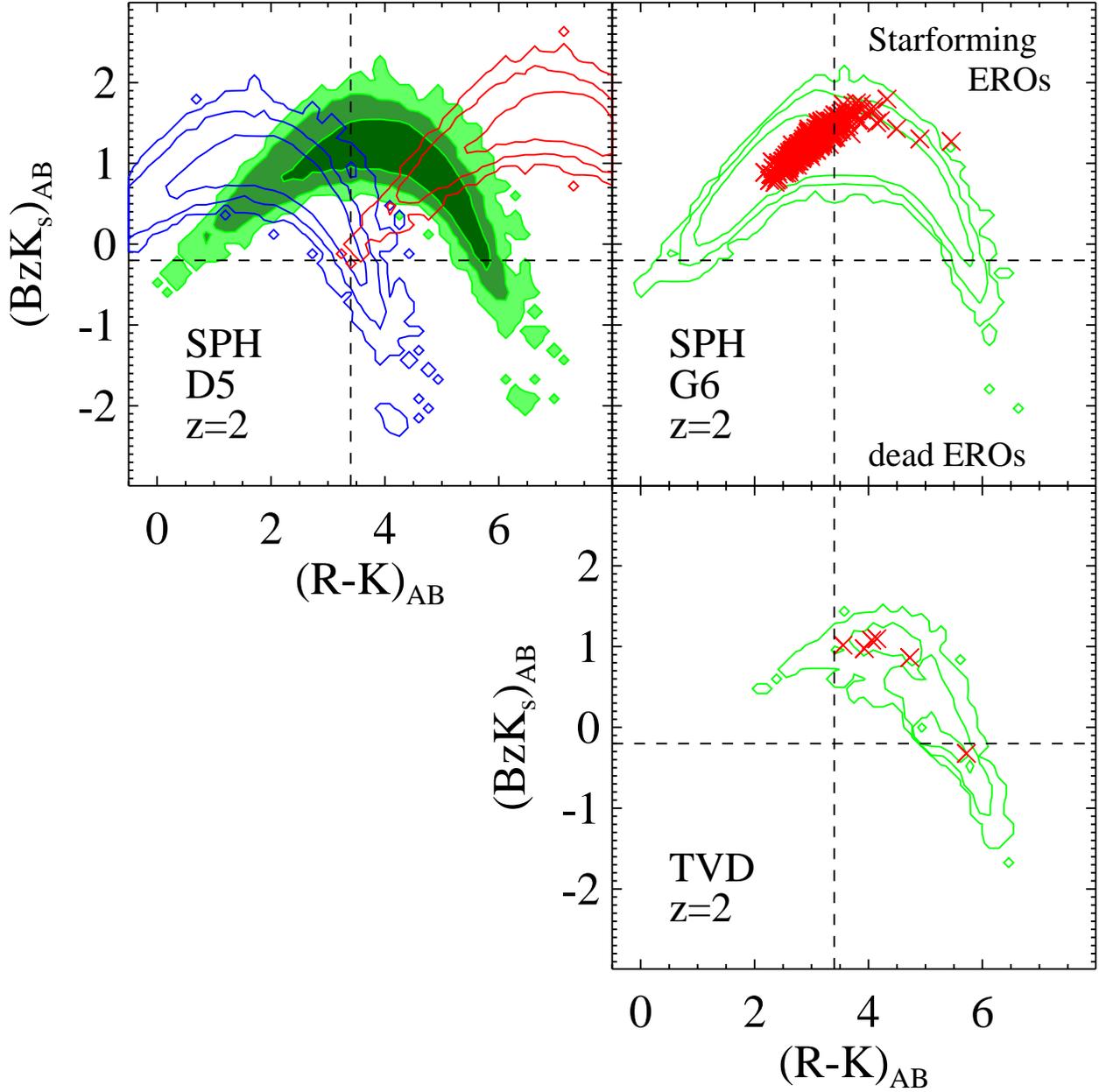}
\caption{
$BzK_s$ versus $R-K_s$ diagram of simulated galaxies at $z=2$, both for
SPH and TVD runs. Similar to Figure~\ref{f5.eps},  in the top left 
panel for the SPH D5 runs, three different sets of distributions are shown, 
corresponding to $E(B-V)=0.0$ (black), 0.4 (blue), and 1.0 (red). 
In the panels for the SPH G6 and TVD run, only the case of $E(B-V)=0.4$
is shown. The red crosses overplotted in the SPH G6 and the TVD
panels are the galaxies that are brighter than $K_s=22$
(i.e., $K_{s,{\rm vega}}<20$).}
\label{f6.eps}
\end{figure}

Figure~\ref{f6.eps} shows $BzK_s$ versus $R-K_s$ diagram of simulated  
galaxies at $z=2$ for both SPH and TVD runs. Similar to 
Figure~\ref{f5.eps}, in the top left panel for the SPH D5 run, we show three 
distributions corresponding to extinction $E(B-V)=0.0$ (black), 0.4 (blue), 
and 1.0 (red). In the panels for the SPH G6 and TVD runs, only the case for 
$E(B-V)=0.4$ is shown. The red crosses overplotted in the SPH G6 and the 
TVD panels are the galaxies that are brighter than $K_s=22$ (i.e., 
$K_{s,{\rm vega}}< 20$).  This figure corresponds to Fig.~15 in 
\citet{Daddi04b}, in which they showed that about 50\% of $BzK_s$ 
($>-0.2$) selected galaxies at $z>1.4$ have ERO colors ($R-K_s>3.4$).  
In the SPH G6 run, 13\% of galaxies at $z=2$ with $BzK_s>-0.2$
and $K_s<22$ satisfy the ERO color criteria, and the corresponding  
number density is $5.6\times 10^{-5}h^3\mpc^{-3}$ for the star-forming EROs.
In the TVD run, all galaxies brighter than $K_s=22$ satisfy the ERO color  
criteria for $E(B-V)=0.4$, and the corresponding number density is 
$5.6\times 10^{-4}h^3\mpc^{-3}$.


\subsection{Star formation histories of EROs}
\label{sec:sfr}

Figure~\ref{f7.eps} shows the star formation histories of the reddest 
EROs that satisfy $K_{s,AB}<22$ (i.e., $M_\star\gtsim 1\times 10^{11} 
\himsun$) with a bin-size of 10 Myrs. The top two panels show two galaxies 
from G6 run z=2 output, and the bottom two panels show two galaxies from 
the TVD z=2 output (note the logarithmic scale of the ordinate). 
This is a composite star formation history of all the 
progenitors that end up in the galaxy with properties indicated in the 
legend at $z=2$, therefore the early star formation may be attributed 
to several progenitors. 

Panel (c) is the one in the TVD run that satisfies the $BzK<-0.2$ criterion 
as well as the magnitude cut, formally being the only `dead, red, \& old' 
massive ERO with $\Mstar > 1 \times 10^{11}\himsun$ according to the 
$BzK$ criteria. Indeed, this galaxy has built up most of its stellar mass 
by $z=3$, and star formation is almost absent between $z=3$ and $z=2$; 
i.e., passively evolving. On the other hand, the galaxy shown in panel 
(d) has little star formation in-between $z=5$ and $z=3$, but somewhat
significant star formation in-between $z=3$ and $z=2$.  
This history allows it to pass the ERO criterion, but not the $BzK<-0.2$ 
criterion, and it does not qualify as a `dead, red, \& old' 
massive ERO. 

The two EROs from the SPH G6 run shown in panels (a) and (b) experience
a moderate peak of star formation in-between $z=5$ and $z=3$, but the 
star formation does not die away in-between $z=3$ and $z=2$, making 
it slightly too blue to satisfy the $BzK<-0.2$ criterion.


\section{Discussion \& Conclusions}
\label{sec:discussion}

We have used two different types of hydrodynamic cosmological simulations
(Eulerian TVD and Lagrangian SPH) to study the properties of massive  
galaxies at $z=1-3$ in a $\Lam$CDM universe. Our emphasis has been on the  
stellar mass and near-IR colors of galaxies, and on a comparison of our 
results with observations, including those of the GDDS, {\it K20}, FIRES, 
and GOODS surveys.

Our results suggest that hydrodynamic simulations based on the $\Lam$CDM 
model do not exhibit the 
``{\it mass-scale problem}''; i.e.~there is no obvious difficulty for the
simulations to produce a sufficiently large number density of massive  
galaxies at high redshift, unlike some of the semi-analytic models.
In particular, we find that the magnitude 
limit of $K_s=22$ (i.e., $K_{s, {\rm vega}}\simeq 20$ mag) of the {\it K20} 
survey roughly corresponds to the stellar mass $\Mstar \sim 
10^{11}\himsun$ at $z=2$, and the number density above the magnitude 
limit is $\sim (3-6) \times 10^{-4}\,h^3\,\mpc^{-3}$ at $z=2$. The 
simulated number density above the stellar mass limit of $\Mstar > 
10^{11}\himsun$ agrees very well with these values. Our simulations can 
therefore account for the observed number density of massive galaxies 
found by the {\it K20} survey.

The mean stellar mass of UV selected galaxies with $R<25.5$ at $z=2$ by
\citet{Steidel04} is $\avg{\log \Mstar}\sim 10.3$, and their space  
density is $\sim 6\times 10^{-3}\,h^3\,\mpc^{-3}$ \citep{Erb04}.  
Therefore the mean stellar mass of the current UV selected sample is 
smaller by an order of magnitude than the {\it K20} galaxies, while the 
number density is higher by an order of magnitude.  Roughly 
$\lesssim 10$\% of the UV selected sample has $\Kvega<20$, 
$\Mstar>10^{11}\Msun$, $(J-K)_{\rm vega}>2.3$ \citep{Erb04}, so currently 
the overlap between the {\it K20} galaxies and the UV-selected galaxies 
by \citet{Steidel04} seems to be at most 10\% of the UV selected
sample at $z\sim 2$. Of course, the UV selection will miss the reddest
galaxies in the {\it K20} sample.  When the magnitude limit of the  
$K$-band selected sample is brought down to $K_s\sim 24$ (i.e., 
$\Kvega=22$, as already for the FIRES data), the $K$-band selected 
sample will contain galaxies with $\Mstar \gtrsim 10^{10}\,\Msun$ at 
$z\sim 2$, and the space density of the two samples (UV selected and 
$K$-band selected) will be comparable at 
$\sim 6\times 10^{-3}\,h^3\,\mpc^{-3}$. At that point, the overlap between 
the two samples may increase to a fraction higher than 10\%.  In 
\citet{Nachos2}, we showed that the total number density of galaxies 
with $R<25.5$ in our simulations was about 
$2\times 10^{-2}\,h^3\,\mpc^{-3}$ when a uniform extinction of 
$E(B-V)=0.15$ was assumed.  This is twice the value of Erb et
al.'s estimate, therefore we can obtain a consistent picture if the other 
half of the galaxies have higher extinction values $E(B-V)\gtrsim 0.4$ 
and can only be detected in the $K$-band survey.

The answer to the ``{\it redness problem}'' remains somewhat uncertain  
owing to the unknown dust content of the EROs and the number fraction of  
dusty star-forming galaxies within the observed ERO samples. We have 
studied the $I-K$, $R-K_s$, and $J_s-K_S$ colors of simulated galaxies 
by applying uniform extinction values to the entire samples. In all 
cases, our simulations suggest that the simulated EROs need to have 
extinction values greater than $E(B-V)=0.4$ in order to have colors as 
red as $(I-K)_{\rm vega}>4$, $(R-K_s)_{\rm vega}>5$, or 
$(J_s-K_s)_{\rm vega}>2.3$.  This is consistent with the extinction values 
estimated for the EROs by observational means (e.g., Cimatti et al. 
2002a; Daddi et al. 2004a; van Dokkum et al. 2004; F\"{o}rster Schreiber 
et al. 2004).

The number density of EROs and DRGs (Distant Red Galaxies, see 
Section~\ref{sec:observations}) in our simulations for a uniform
extinction of $E(B-V)=0.4$ is summarized in Table~\ref{table:ERO}.  A  
robust comparison between theoretical models and observations is currently  
limited by our poor knowledge of the amount of dust in EROs and the 
number fraction of highly obscured starburst galaxies in the ERO samples. 
However, the numbers listed in 
Table~\ref{table:ERO} should serve as a benchmark for more stringent 
comparisons in the future. If the median extinction of EROs is close to 
$E(B-V)\sim 0.4$ as suggested by \citet{Daddi04b}, then the numbers 
summarized in Table~\ref{table:ERO} may not be so far from the actual 
values.  This speculation is perhaps not so unreasonable as it may seem, 
because the recent morphological studies of EROs find that the fraction 
of early and late type is comparable at $30-40\%$ 
\citep[e.g.,][]{Cimatti03, Cimatti04, Mous04, Cassata05}.  In this case, 
one may plausibly speculate that roughly half of the EROs have extinction  
smaller than $E(B-V)=0.4$, and the other half has $E(B-V)>0.4$.

As we described in Section~\ref{sec:rk}, our simulations are able to  
account for the comoving number density of EROs of a few $\times
10^{-3}\,h^3\,\mpc^{-3}$ at $z=1$ and a few $\times 10^{-4}\,h^3\,\mpc^{-3}$
at $z=2$, provided we assume a uniform extinction of $E(B-V)=0.4$. These
values are comparable to, or even slightly higher than the observational
estimates \citep{Cimatti02a, Mous04, Vaisanen04, Daddi04b}. Taking this  
result at face value, our simulations do not appear to exhibit the 
``{\it redness problem}'' either. However, as discussed in 
Section~\ref{sec:bzk}, if the observed number density of quiescent old 
EROs (that satisfy the criteria $K_s<22$, $BzK_s<-0.2$, and $z-K_s>2.5$) 
is really $\sim 1\times 10^{-4}\,h^3\mpc^{-3}$, then the current SPH 
simulations underpredict the space density of such a population at $z=2$. 
The TVD simulation contained one such object, yielding the correct space 
density, but the statistical uncertainty of this result in a 
$\Lbox=22\himpc$ box is, of course, very large.

Since the space density of massive EROs is fairly low, large simulation
boxsizes with $\Lbox \gtrsim 100\himpc$ are desired for future comparisons
with observations. The controversial question is not just whether the
hierarchical models can account for the space density of EROs, but it  
is the number density of {\it quiescent, old, passively evolving} EROs. 
It appears unlikely that our simulations have a problem in reproducing 
the space density of star-forming EROs. The $BzK_s$ diagram proposed by 
\citet{Daddi04b} provides a useful test for this interesting population 
of old EROs which would otherwise be difficult to separate out clearly 
owing to dust confusion. Indeed, a recent study by \citet{Daddi05} 
using the Hubble Ultra Deep Field data shows that the bulk of objects 
selected by $K_s<22$, $BzK<-0.2$, and $z-K>2.5$ are old and passive 
early types at redshifts $1.4<z<2.0$, with space density of 
$\sim 10^{-4}\,\mpc^{-4}$.  The ``redness problem'' is therefore
perhaps better described as a ``dead massive ERO problem''.

In this regard the connection between EROs and AGNs is intriguing.  On  
the observational side, it is shown by {\it Chandra} and {\it XMM-Newton}  
that a sizable fraction ($\gtrsim 15$\%) of the $2-10$ keV selected 
sources are associated with EROs \citep{Rosati02, Mainieri02}. The 
majority of the X-ray emitting EROs studied so far strongly suggests that 
the bulk of this population is composed of obscured AGNs, at least for 
the brightest X-ray fluxes \citep[e.g.,][]{Alexander02, Alexander03, 
Sever05}.  On the theoretical side, \citet*{Springel04} recently 
investigated the effect of AGN feedback during major mergers of gas-rich 
spiral galaxies using hydrodynamic simulations. 
\citet*{Springel05} showed that mergers with black holes produce 
ellipticals that redden much faster owing to the 
suppression of further star formation by a strong outflow generated by 
the central black hole \citep[see also][]{Sazonov05, Scan04}, when
feedback effects are normalized to reproduce the observed correlation 
between black hole mass and stellar velocity dispersion \citep*{DiM05}. 
Therefore, AGN feedback may play a critical role in producing red, old, 
and passively evolving massive EROs. This scenario can be tested with 
future cosmological simulations which self-consistently include black 
hole growth and AGN feedback processes.

Although the different simulation methods analyzed here broadly agree  
on the properties of high redshift galaxies, there are also interesting  
systematic differences between them. In general, the TVD run tends to 
predict a somewhat higher number density of EROs at all redshifts 
compared with the SPH simulations. This may owe to the fact that the 
star formation history in the TVD simulation is more sporadic than that 
of the SPH simulations, as we showed in \citet{Nachos2}.  Therefore at 
any given time, the stellar population of the simulated galaxies in the 
TVD run has more time to become redder between starbursts, without 
being so frequently affected by the blue light from the most recent star 
formation.  This difference in the nature of the star formation history 
probably owes to a combination of differences in the details of the 
parameterization of the star formation physics, the strength of the 
feedback effects, the hydrodynamic methods, and the numerical 
resolution reached in the different simulations. 

\begin{figure}
\epsscale{1.0}
\plotone{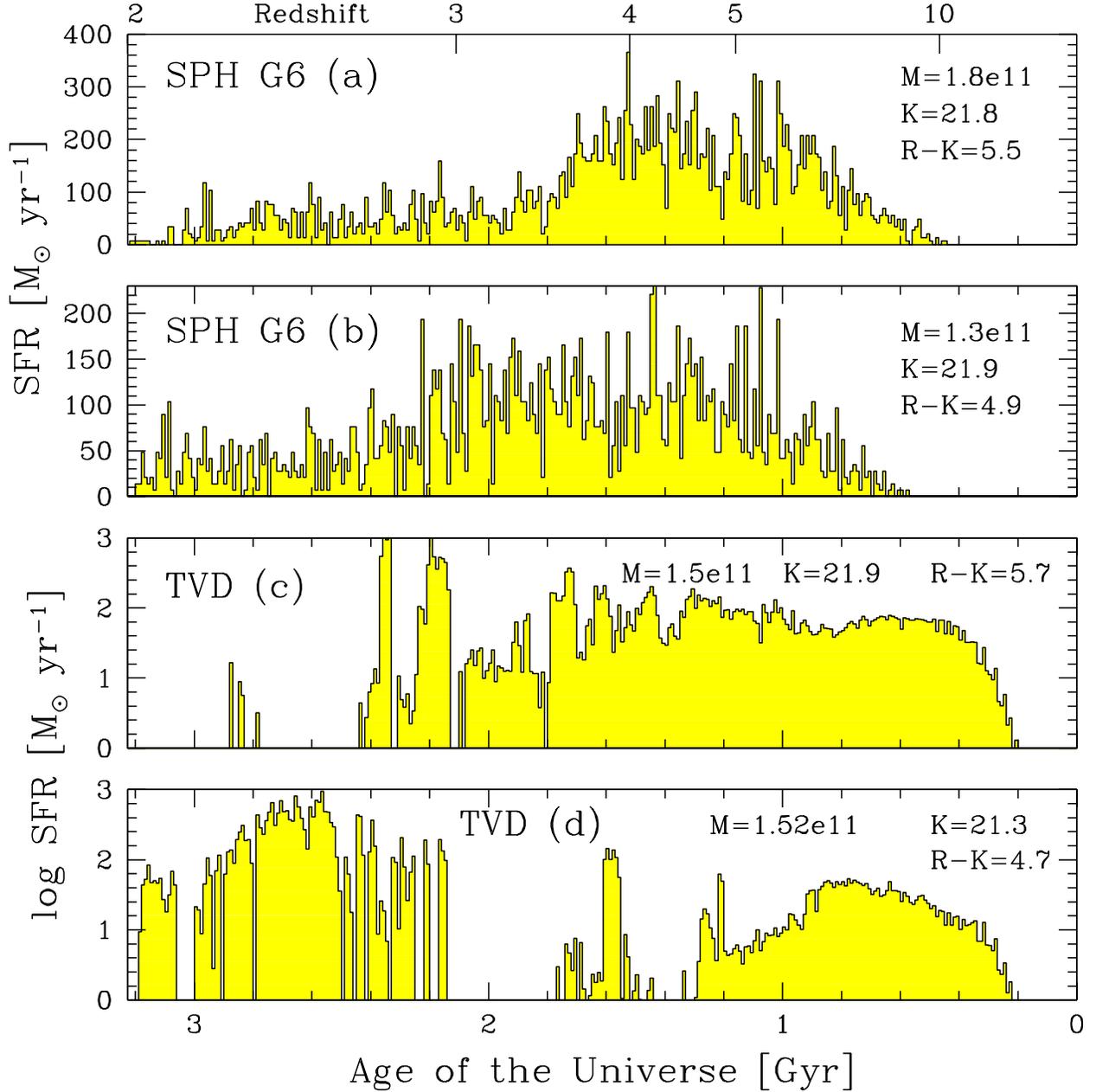}
\caption{Star formation histories of the reddest EROs that satisfy
 $K_{s,AB}<22$ (i.e., $M_\star\gtsim 1\times 10^{11} \himsun$) in our 
simulations. 
The top two panels show two galaxies from the G6 run z=2 output, and the 
bottom two panels show two galaxies from the TVD z=2 output (note the
logarithmic scale of the ordinate). 
Panel (c) is the one in TVD run that satisfies the $BzK<-0.2$ criteria 
as well, being the only `dead \& red' massive ERO with $\Mstar > 
1 \times 10^{11}\himsun$. Note that this is a composite star formation 
history of all the progenitors that end up in the galaxy with properties 
indicated in the legend at $z=2$, therefore the early star formation 
may be attributed to several progenitors. }
\label{f7.eps}
\end{figure}

Finally, we comment on the differences between our hydrodynamic
simulations and semi-analytic models of galaxy formation.
As described in Section~\ref{section:intro} and by \citet{Nachos1},
our simulations have a star formation rate history that peaks at
much higher redshift ($z\ge 5$) than that of most published 
semi-analytic models. We argue that this is one of the reasons why 
our hydrodynamic simulations are more successful in reproducing the 
properties of red massive galaxies at $z\sim 2$. Apparently, the star 
formation rate at high redshift ($z\ge 3$) is strongly suppressed by 
supernovae feedback in these semi-analytic models. While this helps 
them to match the faint-end of the galaxy luminosity function at 
$z=0$, they have problems in explaining EROs at high redshift. 
However, we note that a more recent revised version of the 
semi-analytic model by \citet{Baugh05} enhances the burst-mode of 
star formation and obtains better agreement with the infra-red 
galaxy number counts and the redshift distribution of submillimeter 
galaxies at high redshift. Note that both TVD and SPH simulations 
(see Figure~\ref{f7.eps}) automatically includes high amplitude 
bursts in response to infalling lumps of gas. 
Another revised semi-analytic model by \citet{Granato04} incorporates 
the effects of AGN feedback, and now succeeds in producing red massive 
spheroidal galaxies at early times, also achieving better agreement 
with the observed infra-red to submillimeter galaxy counts 
\citep{Silva05}. The results on the star formation rate history 
obtained by these recent semi-analytic models seem to be much closer 
to our hydrodynamic simulations and show more intense star formation 
at high redshifts. It would be interesting to see whether these 
revised semi-analytic models have the ``{\it dead massive ERO problem}'' 
by examining the $BzK$ diagrams of massive galaxies at $z\sim 2$.


\acknowledgments

This work was supported in part by NSF grants ACI 96-19019,
AST 00-71019, AST 02-06299, and AST 03-07690, and
NASA ATP grants NAG5-12140, NAG5-13292, and NAG5-13381.
The SPH simulations were performed at the Center for Parallel
Astrophysical Computing at Harvard-Smithsonian Center for
Astrophysics. The TVD simulations were performed at the National
Center for Supercomputing Applications (NCSA).



\begin{deluxetable}{cccccc}
\tablecolumns{6}
\tablewidth{0pc}
\tablecaption{Simulations}
\tablehead{
\colhead{Run} & \colhead{$\Lbox$ [$\himpc$]} & \colhead{$N_{\rm 
mesh/ptcl}$} & \colhead{$m_{\rm DM}$ [$\himsun$]}  & \colhead{$m_{\rm  
gas}$ [$\himsun$]} & \colhead{$\Del\ell$ [$\hikpc$]}
}
\startdata
TVD: N864L22$^a$ & 22.0 & $864^3$ & $8.9\times 10^6$ & $2.2\times 10^5$  
& 25.5\cr
SPH: D5$^b$ & 33.75 & $324^3$ & $8.2\times 10^7$ & $1.3\times 10^7$ &  
4.2\cr
SPH: G6$^b$ & 100.0 & $486^3$ & $6.3\times 10^8$ & $9.7\times 10^7$ &  
5.3\cr
\enddata
\tablecomments{
Parameters of the primary simulations on which this study is based.
The quantities listed are as follows: $\Lbox$ is the simulation
box size, $N_{\rm mesh/ptcl}$ is the number of the hydrodynamic mesh
points for TVD, or the number of gas particles for SPH, $m_{\rm DM}$
is the dark matter particle mass, $m_{\rm gas}$ is the mass of the
baryonic fluid elements in a grid cell for TVD, or the masses of the
gas particles in the SPH simulations. Note that TVD uses $432^3$
dark matter particles for N864 runs. $\Del\ell$ is the size of the
resolution element (cell size in TVD and gravitational softening
length in SPH in comoving coordinates; for proper distances, divide
by $1+z$). The upper indices on the run names correspond to the
following sets of cosmological parameters:
$(\Om, \Ol, \Ob, h, n, \sigma_8) =
(0.29, 0.71, 0.047, 0.7, 1.0, 0.85)$ for (a), and
$(0.3, 0.7, 0.04, 0.7, 1.0, 0.9)$ for (b).
}
\label{table:simulation}
\end{deluxetable}


\begin{deluxetable}{cccccccccc}
\tabletypesize{\scriptsize}
\tablecolumns{9}
\tablewidth{0pc}
\tablecaption{Number density of EROs and DRGs in the simulations}
\tablehead{
\colhead{} &
\multicolumn{2}{c}{$z=1$} & \colhead{} &
\multicolumn{3}{c}{$z=2$} & \colhead{} &
\multicolumn{2}{c}{$z=3$} \\
\cline{2-3} \cline{5-7} \cline{9-10}
\colhead{} & \colhead{$(I-K)$\tablenotemark{a}} &  
\colhead{$(R-K_s)$\tablenotemark{b}} & \colhead{} &  
\colhead{$(R-K_s)$\tablenotemark{b}} &  
\colhead{$(J_s-K_s)$\tablenotemark{c}} &  
\colhead{$(z-K_s)$\tablenotemark{d}} & \colhead{} &  
\colhead{$(R-K_s)$\tablenotemark{b}} &  
\colhead{$(J_s-K_s)$\tablenotemark{c}}}
\startdata
SPH & $(2-3)\times 10^{-4}$ & $3.2\times 10^{-3}$ & & $5.6\times  
10^{-5}$ & $1.4\times 10^{-3}$ & $2.2\times 10^{-4}$ &  & $5.0\times  
10^{-6}$ & $1.3\times 10^{-3}$\\
TVD & $2.5\times 10^{-3}$ & $4.2\times 10^{-3}$ &  & $5.6\times  
10^{-4}$ & $7.4\times 10^{-3}$ & $5.6\times 10^{-4}$ & & --- &  
$1.5\times 10^{-3}$
\enddata
\tablenotetext{a}{$I-K>2.6$ and $I<25$, for comparison with the GDDS  
data}
\tablenotetext{b}{$R-K_s>3.4$ and $K_s<22$, for comparison with the  
GOODS data}
\tablenotetext{c}{$J_s-K_s>1.34$ and $K_s<24.4$, for comparison with  
the FIRES data on DRGs}
\tablenotetext{d}{$z-K_s>2.5$ and $K_s<22$, for comparison with the  
{\it K20} data}
\tablecomments{
Number density of EROs and DRGs in units of $h^3\,\mpc^{-3}$ in our 
simulations satisfying the above color-selection and magnitude limit, 
assuming a uniform extinction of $E(B-V)=0.4$ for all galaxies.
The density for the $R-K_s$ selection at $z=3$ is clearly affected by  
the limited boxsize, and is not very reliable. The number density for 
the $J_s-K_s$ selection is higher than other cases because of the deeper 
magnitude limit (corresponding to the FIRES data). }
\label{table:ERO}
\end{deluxetable}


\end{document}